# Compositional Effects on Structure, Dynamics, Thermodynamic and Mechanical Properties of Zr-Cu-Al alloys

**Kamal G. Soni,*  Jayraj P. Anadani, Mitanshu B. Vahiya, Kirit N. Lad***

*Department of Physics, Sardar Patel University, Vallabh Vidyanagar-388120, Gujarat, INDIA*

## Abstract

Zr-Cu-Al alloys belong to a commercially important family of CuZr-based alloys that form bulk metallic glasses on microalloying of Al. However, the identification of compositions with good glass-forming ability and desirable properties from a vast compositional space remains a major challenge due to complex compositional effects on the structure, dynamics and properties. In the present work, we report molecular dynamic investigations of structure, dynamics, thermodynamic and mechanical properties of $Zr_{50}Cu_{50-x}Al_x$ and $Cu_{50}Zr_{50-x}Al_x$ alloys (x=5,10,15,20,25,30,40) covering a wide compositional space. Our results and findings lead to some important conclusions that could serve as overarching guidelines for choosing good glass-forming alloy compositions that give Zr-Cu-Al glasses with tailored thermal and mechanical properties. Overall, present results suggest that a good glass-forming Zr-Cu-Al alloy composition leading to an MG with good thermal and mechanical properties should be Cu-rich with Zr concentration in the window 30% – 35% and Al% ≤ 20. Our results also highlight the impact of icosahedral short- and medium-range ordering on the dynamics and mechanical properties of the alloys. It is observed that the fractions of the full icosahedra (<0,0,12,0>) and the degree of their interconnectivity are directly correlated to the structural relaxation, diffusion, dynamic heterogeneity and mechanical properties.

---

* Corresponding authors: knlad-phy@spuvvn.edu, kamalsoni@spuvvn.edu

## 1. Introduction

Zr-Cu-Al alloys belong to the family of CuZr-based alloys that form bulk metallic glasses (BMG) and possess high mechanical strength, making them potential candidates for high-tech structural applications. [1–10]    Despite being the simplest BMG-forming multicomponent system, its glass-forming ability (GFA) and mechanical properties exhibit a sensitive compositional dependence, [2,7,8,11] and finding compositions with high GFA and desirable mechanical properties remains a long-standing challenge. To reduce the time and cost of searching for the best glass-forming compositions through the trial-and-error approach, semi-empirical and hybrid approaches combining modelling, experiments and simulations, guided by structural and thermodynamic factors such as the atomic sizes, enthalpy of mixing, have been employed to pinpoint compositions of Zr-Cu-Al alloys with high GFA. [12–17] Understanding the role of structural and thermodynamic factors in governing GFA and the properties of these alloys has been an area of intense research for the last two and half decades, where the topological short- and medium-ranger ordering, chemical ordering and their effect on the dynamics, GFA and properties have been the major focus of the investigations. [18–40] Advances in experimental synthesis and characterization techniques, high-performance computational resources and methods  have led to the development of various impressive combinatorial  [41–44] and data-driven methods [45] involving high throughput synthesis and characterization techniques that can identify the best glass-forming composition from a vast compositional space of the multicomponent metallic alloys. Data-driven machine learning methods [46–50] have also evolved to design and discover metallic glasses with desired physical properties by exploring the structure-property landscape of glass-forming metallic alloys. Although the high throughput experimental methods are successful, it is still overwhelmingly challenging to explore the entire composition space of the metallic glasses. The success of the data-driven methods, on the other hand, depends on the availability of large and reliable data on the structure and properties of metallic alloys. Thus, a thorough knowledge of the compositional effects on the structure, dynamics and properties is essential for reliable prediction of alloy compositions with high GFA and the resultant MGs with optimum properties.

In the present work, we undertake molecular dynamic investigations of structure-property correlations in Zr-Cu-Al alloys covering a wide compositional space. The microalloying approach adopted in most of the previous studies usually amounts to the random substitution of Zr and Cu atoms by Al atoms in a given proportion. For example, on minor addition of Al in proportion $y$ in a binary metallic alloy, $Zr_xCu_{100-x}$, Zr and Cu atoms are randomly substituted by Al atoms, giving a ternary alloy $(Zr_xCu_{100-x})_{100-y}Al_y$. Such substitution of Zr and Cu atoms obscures the understanding of the compositional effects on the glass formation in the resultant ternary alloys due to competing atomic interactions and topological factors. In this sense, studies that give a comprehensive picture of the compositional effects in Zr-Cu-Al alloys covering a broad composition range are still scarce. To get clearer insights into the complex interplay of the chemical and topological ordering factors on the structure, dynamics, thermodynamics and mechanical properties of metallic alloys, it is rather more instructive to substitute Zr (or Cu) by Al in a given proportion while keeping the proportion of Cu fixed and the other way round. Therefore, we investigate $Zr_{50}Cu_{50-x}Al_x$ (ZCA) and $Cu_{50}Zr_{50-x}Al_x$ (CZA) $(0 \leq x \leq 40)$ alloys to understand how the substitution of Cu and Zr by Al in ZCA and CZA, respectively impacts their structure, dynamics and properties. We thoroughly examine the changes in the average structure, topological short- and medium-range order, chemical ordering, structural relaxation, diffusion processes, and mechanical properties (Young's modulus and yield stress) due to increasing Al concentration in the ternary alloys. Our results show distinctly different trends in the variation of the structural, dynamical and thermodynamic properties with Al concentration in ZCA and CZA alloys. While the increasing substitution of Cu by Al does not lead to a significant change in the structure and dynamics of ZCA alloys, the reduction of Zr concentration on its substitution by Al in CZA alloys results in remarkable and non-monotonic changes in the structure and dynamics. The comprehensive picture of the compositional dependence of the structural and dynamical variables, thermodynamic and mechanical properties would serve as an important guide for the synthesis of Zr-rich or Cu-rich Zr-Cu-Al MGs with optimized microstructure, thermal stability and mechanical properties.

## 2. Simulation Details and Methods

Classical molecular dynamics (MD) simulations of $Zr_{50}Cu_{50-x}Al_x$ and $Zr_{50}Cu_{50-x}Al_x$ (x = 5, 10,15,20,25,30,40) alloys are performed using LAMMPS package. [51] A cubic simulation box with a total of 13500 atoms with the number of Zr, Cu and Al atoms corresponding to a given alloy composition was considered and the periodic boundary conditions were employed in all three directions. For example, in case of $Zr_{50}Cu_{45}Al_5$, $N_{Zr} = 6750$, $N_{Cu} = 6075$ and $N_{Al} = 675$. A many-body embedded-atom model (EAM) potential, [22] which is extensively used for MD simulation of Zr-Cu-Al alloys, is used. Starting with an initial FCC configuration of a system at 2000 K, a homogenized melt of the alloy is generated by running a simulation in isothermal-isobaric (NPT) ensemble for a sufficiently long simulation time of the order of 2 ns. All the simulations were performed at a constant pressure of 0 bar. Nose-Hoover thermostat was used to control the temperature. The leapfrog Verlet algorithm was used to integrate the equations of motion and update the positions and velocities of the atoms at every time step of 2 fs, which is small enough to accurately minimize fluctuations in the total energy. The homogenized liquid configuration of the alloy at 2000 K is cooled down to 300 K at the rate of 0.1 K/ps. The alloy configurations at intermediate temperatures during the quench run were recorded for further simulations and analysis.

The average structure of the amorphous alloys has been analysed in terms of the pair correlation function, g(r), and the static structure factor, S(q) given by

$$S(q) = 1 + 4\pi\rho \int_0^\infty r^2 \frac{sin(qr)}{qr} [g(r) - 1]\, dr \qquad (1)$$

where $\rho$ is number density.

To study the chemical short-range order (CSRO), the Warren-Cowley CSRO parameter is computed using the definition [52]

$$\alpha_{ij} = 1 - \frac{N_{ij}}{c_j N_i} \qquad (2)$$

where $c_j$ denotes the concentration of element $j$ in the alloy. Negative values of $\alpha_{ij}$ indicate the preference of atom of type $i$ to have atoms of type $j$ as their first neighbors. Positive $\alpha_{ij}$ is the sign of avoidance of atoms of type $j$ by the atoms of type $i$.

Local structure is analyzed using the Voronoi tessellation method through Voro++ software package.[53] Starting with the quenched liquid configuration at 300 K, a production run in NPT ensemble was performed to get 25 different configurations. These configurations were subsequently subjected to an energy minimization process using the conjugate gradient algorithm to bring the system to a local minimum on the potential energy landscape (PEL). The fundamental information extracted in the form of Voronoi indices, $\langle n_3, n_4, n_5, n_6 \rangle$, for different geometrical polyhedra is used to analyze the short & medium-range order in the studied alloys.

## 3. Results and Discussion

### 3.1 Glass transition, enthalpy of chemical mixing, and specific heat

To get the first-hand information of the compositional effects on the glass formation in the ZCA and CZA systems, we look at the fictive temperature ($T_f$) – a temperature at which an inflection is observed in the potential energy (or enthalpy or volume) vs. temperature curve during the rapid cooling of a liquid as seen in Fig. S1 of Supplementary Materials. As the glass transition is a kinetic phenomenon that depends on the cooling rate, $T_f$ would correspond to the glass transition temperature, $T_g$, at the critical cooling rate, which is necessary to ascertain glass formation in the given system. Fig. 1 shows the variation in the $T_f$ and the average potential energy at $T_f$ with the Al% in the ZCA and CZA alloys. The potential energy shows a small variation over the range of Al%, $5 \leq x \leq 40$, and $T_f$ is almost insensitive to the Al concentration in the ZCA alloy. In CZA alloys, the potential energy and $T_f$ show remarkable variation with the Al%. $T_f$, which is 762 K for 5% Al, decreases to 625 K for 40% Al in CZA alloy. From the PEL perspective, these results give useful insights into the compositional effects on the dynamics of the supercooled liquids, glass transition and thermodynamics. The remarkable increase in the average potential energy at $T_f$ on increasing substitution of Zr by Al in CZA alloys indicates a reduction in the roughness of the PEL, i.e. a decrease in the depths of local minima (meta-basins) on the PEL. The PEL roughness governs the system dynamics in the supercooled region. [54] As the system explores deeper PEL minima with decreasing temperature, the system dynamics slow down rapidly, making the inter-basin(minima) transitions difficult until the glass transition where translation degrees of freedom are frozen, and the system is trapped in one of the minima. So, the PEL roughness reduction implies that the system dynamics would slow down less rapidly, and the glass transition would occur at a lower temperature, unlike the case of PEL with high roughness. Thus, the observed decrease in $T_f$ in tandem with the increase in the average potential energy suggests significant changes in the PEL with increasing Al% in the CZA alloys. For the ZCA alloys, nearly the same $T_f$ and a significantly small variation in the average potential energy at $T_f$ suggest a very little change in the PEL and the dynamics due to the compositional changes on substitution of Cu by Al.

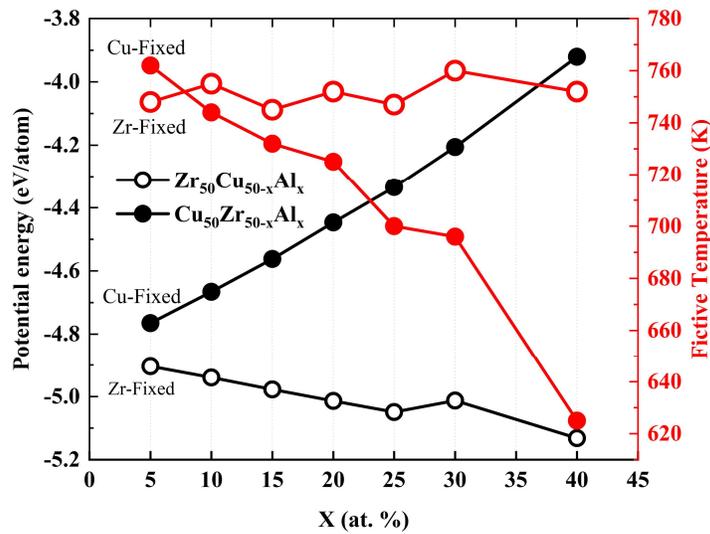

**Figure 1**: Fictive temperature $T_f$ and the corresponding average potential energy for different Al% in ZCA and CZA alloys. $T_f$ is determined from the potential energy vs. temperature curve during rapid cooling in NPT ensemble at the same cooling rate for all the compositions.

Two distinct scenarios observed in the compositional effects of the substitution of Cu by Al and Zr by Al in the ZCA and CZA alloys, respectively, clearly point towards the significance of the competing chemical interactions among the constituent atoms of the alloys. To gain preliminary insights into the compositional impact of these interactions on the thermodynamics of the alloys, we look into the enthalpy of chemical mixing defined as [55]

$$\Delta \mathrm{H}^{Chem} = 4 \sum_{i \neq j}^{3} \Delta H_{ij}^{mix} c_i c_j \qquad (2)$$

where $\Delta H_{ij}^{mix}$ represents the enthalpy of mixing for $i$ and $j$ elements. $c_i$ and $c_j$ are the molar percentages of the elements $i$ and $j$. The enthalpies of mixing for Zr-Cu, Zr-Al and Cu-Al interactions are -23 kJ/mol, -44 kJ/mol and -1 kJ/mol, respectively. [56] The results in Fig. 2 show that $\Delta \mathrm{H}^{Chem}$ decreases monotonically with increasing Al% in the ZCA alloy, whereas it changes non-monotonically with the Al% in the CZA alloy – an initial small decrease up to Al% = 15 is followed by a nonlinear increase up to Al% = 40.

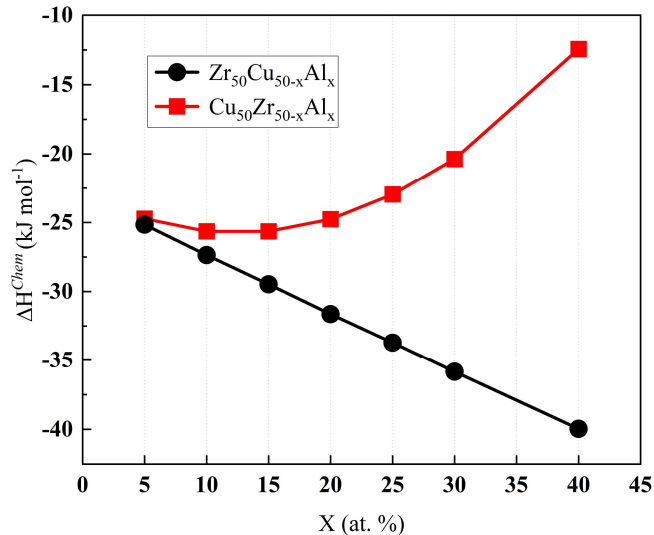

**Figure 2**: Enthalpy of chemical mixing in ZCA and CZA alloys.

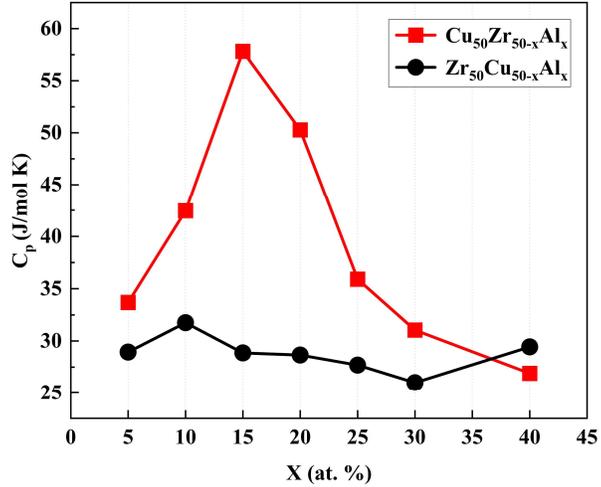

**Figure 3:** Specific heat of ZCA and CZA alloys at 850 K.

We also look into the composition dependence of the specific heat, $C_p$, in ZCA and CZA alloys. $C_p$ for the two alloys at 850 K are derived using its relation with the enthalpy fluctuations in NPT ensemble, $C_p = [\langle H^2 \rangle - \langle H \rangle^2]/k_B T^2$, where the averaging was done over a considerably long (~ns) simulation trajectory. The results are plotted in Fig. 3. While $C_p$ changes remarkably with the Al% in the CZA alloys, the observed variation in the $C_p$ is relatively very small in ZCA alloys. A large increase in $C_p$ up to Al% = 15 and a subsequent fast decrease for Al% > 15 can be observed in CZA alloys.

Such a starkly different composition dependence of $\Delta H^{Chem}$ and $C_p$ highlights the importance of competing chemical interactions among the atoms in ZCA and CZA alloys. Analysis of the chemical short-range order, presented in Sec. 3.3, gives a more elaborate understanding of the change in the interatomic chemical interactions with the compositional changes in the ZCA and CZA alloys.

### 3.2 Structural characteristics in real and Fourier space

The results in the previous section amply indicate significant compositional effects on the average structure of the CZA alloys and less conspicuous changes in the average structure of ZCA alloys. Noting the atomic radii of Zr, Cu and Al to be 1.60 Å, 1.28 Å and 1.43 Å, respectively, the preliminary signatures of the different structural changes in the two alloys can be found from the composition dependence of the number density ($\rho_N$), shown in Fig. 4. For a fixed system size (total number of atoms = 13500) in the MD simulation, the decrease in $\rho_N$ in ZCA alloys and the increase in $\rho_N$ in CZA alloys are on the expected lines from the common hard-sphere atomic perspective. However, it is noteworthy that the degree of the compositional effects on the total $\rho_N$ and $(\rho_N)_{Al}$ after Al% = 15 is larger in CZA alloys than ZCA alloys. The larger changes in the total $\rho_N$ and $(\rho_N)_{Al}$ in CZA alloys can be attributed to the increasing Cu-Al interactions, which have been reported to show significant bond-shortening. [22]

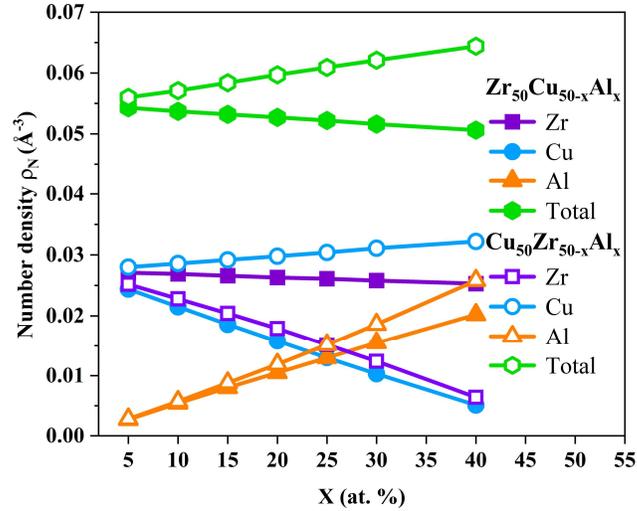

**Figure 4:** Atomic number densities ($\rho_N$) in ZCA and CZA alloys at 850K

Further understanding of the composition dependence of the average structure in the alloys can be gained from the pair correlation function, g(r), at 300K shown in Fig. 5. To acquire clearer structural features in g(r), the inherent structure configurations, extracted as explained in Sec. 2, have been analyzed and the results are depicted in Fig. 5(b). It can be observed that CZA and ZCA alloys exhibit nearly identical structural features up to Al%=15. The common features of the main (first) peak of the total g(r) to be observed are: a pre-peak (or shoulder) corresponding to the Cu-Cu pairs, the central peak for the Cu-Zr pairs, and the subsequent shoulder for the Zr-Zr pairs. In CZA system, the sub-peak corresponding to Cu-Cu pairs becomes the prominent (main peak) for Al% > 20, whereas the main peak corresponding to Cu-Zr pairs is subdued due to the decreasing concentration of Zr atoms. As a result, the main peak position shifts to lower *r*, indicating a decrease in the average first-neighbour distance. Contrary to CZA system, the Cu-Cu pre-peak diminishes in ZCA system for Al% > 20, and the feeble Cu-Zr peak merges with the shoulder peak corresponding to Zr-Zr pairs.

While g(r) gives useful information about the distribution of atoms in real space, the strength of atomic interactions and chemical ordering, its Fourier transform – the static structure factor, S(q) – has been reported to give insights into the degree of structural ordering, interatomic distance and the GFA in metallic glasses through the characteristic features of the first peak – the peak position and its full width at half the maximum (FWHM). [57] Therefore, we obtained S(q) of ZCA and CZA metallic glasses at 300 K, as shown in Fig. 6(a) and Fig. 6(b), respectively. Two opposite trends in the composition dependence of the first peak position

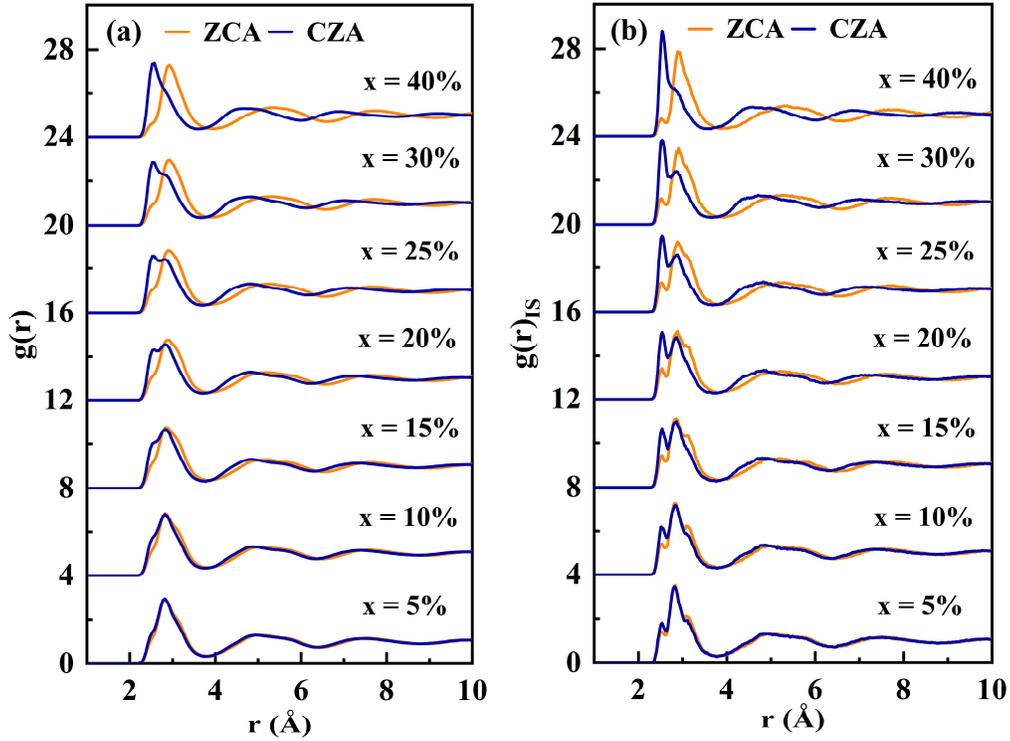

**Figure 5:** Total pair correlation function for ZCA and CZA metallic glasses at 300 K, (a) as-quenched state, (b) from inherent structures obtained through energy minimization

($q_1$) (Fig. 6(d)), maximum value, $S(q_1)$, (Fig. 6(c)) and FWHM ($\Delta q$) (Fig. 6(e)) in ZCA and CZA metallic glasses are observed. According to the well-known Ehrenfest relation [58] between the average interatomic spacing ($d$) and $q_1$, $d \propto \frac{1}{q_1}$, the decreasing $q_1$ with increasing average substitution of Cu by Al in ZCA alloys suggests an increase in $d$. In CZA alloys, where smaller Al atoms substitute the larger Zr atoms, the rise in $q_1$ is a clear manifestation of the decrease in the $d$. The increase in the peak height, $S(q_1)$, in ZCA alloys, indicates an increase in the strength of the average atomic interactions due to increasing Zr-Al interactions. The decrease of $S(q_1)$ in CZA, which suggests weaker average atomic interactions, can be attributed to the decreasing Zr-Al interactions. The FWHM ($\Delta q$) gives more useful information related to the structural correlation and the GFA in the metallic glasses. A larger $\Delta q$ is shown to be associated with a shorter structural correlation, implying a higher degree of disorder and higher GFA. [57] Therefore, the increase in $\Delta q$ with the increase in Al concentration up to 30% (Fig.6(e)) suggests an increasing disorder and GFA in CZA glasses. In contrast, the decrease in $\Delta q$ with Al% (Fig.6(e)) indicates a reduction in the degree of disorder and GFA in ZCA glasses. The correlation between $\Delta q$ and $q_1$ shown in Fig. 6(f) points to a remarkable increase in the GFA up to 30% Al in CZA alloys and a comparatively small increase in the GFA of ZCA alloys. These results are consistent with those reported by Li et al. [45]

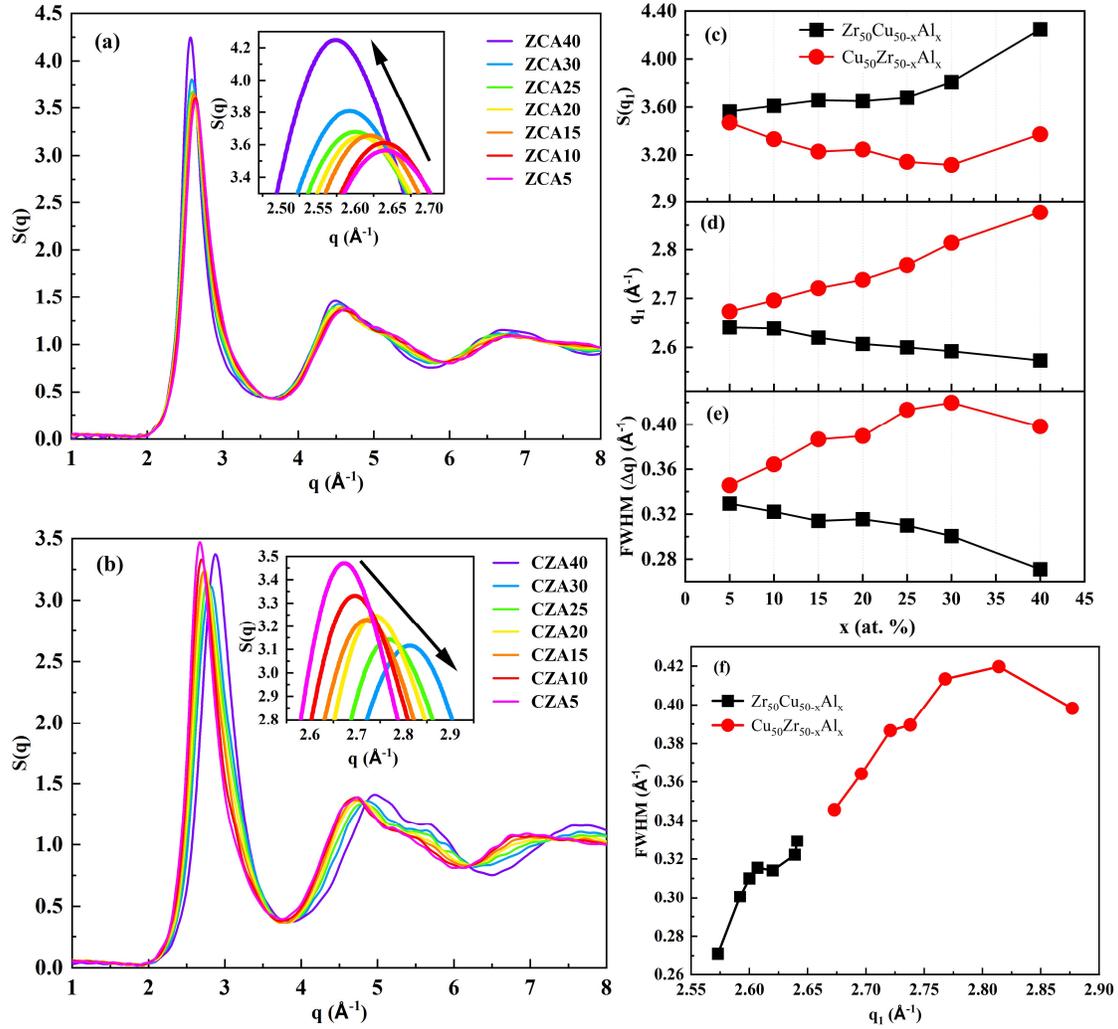

**Figure 6:** Static structure factor of different compositions of metallic glasses: (a) $Zr_{50}Cu_{50-x}Al_x$, (b) $Cu_{50}Zr_{50-x}Al_x$ at 300 K, the insets in the figures give enlarged view of the changes in the first peak position and height; Composition dependence of the first peak characteristics: (c) maximum $S(q_1)$, (d) peak position $q_1$, and (e) FWHM ($\Delta q$); (f) correlation between the peak position ($q_1$) and the FWHM.

A last noteworthy observation from the S(q) results is the splitting of the second peak, which is often linked to the icosahedral short- and medium-range ordering in metallic glasses. [59] While the first sub-peak is evident in both ZCA and CZA glasses, the second sub-peak remains subdued in the former. Moreover, the changes in the second peak are more conspicuous in CZA glasses compared to ZCA. These are the preliminary indicators of a significantly higher degree of icosahedral short- and medium-range ordering in CZA glasses than ZCA glasses.

*3.3 Atomic coordination and chemical short-range ordering*

We analyze the atomic coordination tendencies and the CSRO to get deeper insights into the composition dependence of the average structure of the ZCA and CZA metallic glasses. The distribution of the average atomic coordination number (CN) and the CNs of Zr, Cu, and Al atoms are shown in Fig. 7. The bimodal distributions of the average atomic CN, with one peak centered at 12 and the other at 15, are observed in the two metallic glasses. (Fig. 7(a) & (b)) While a clear bimodal distribution persists up to Al% = 20 in both the glasses, one of the peaks turns into a weaker shoulder beyond that and, appears like a broader unimodal distribution. Nearly equal heights of the two peaks in ZCA metallic glasses suggest strong competition among Cu-Zr, Zr-Al and Cu-Al chemical interactions, and weaker bonding. In the case of CZA, the first peak is significantly more prominent than the second peak. It is an indication of stronger Cu-Zr, and Cu-Al interactions than Zr-Al interactions. The changes in the average atomic CN distribution with the change in Al% are more conspicuous in CZA glasses. This can be attributed to significant changes in the Al neighbour distribution, as observed in Fig. 7(h). The CN distribution curve for Al atoms remains qualitatively similar with a peak at 13 in ZCA glasses (Fig. 7(e)), whereas it shifts towards coordination number 14 in CZA glasses with Al% > 20 (Fig. 7(h)). These results suggest significant changes in CSRO in CZA glasses compared to ZCA.

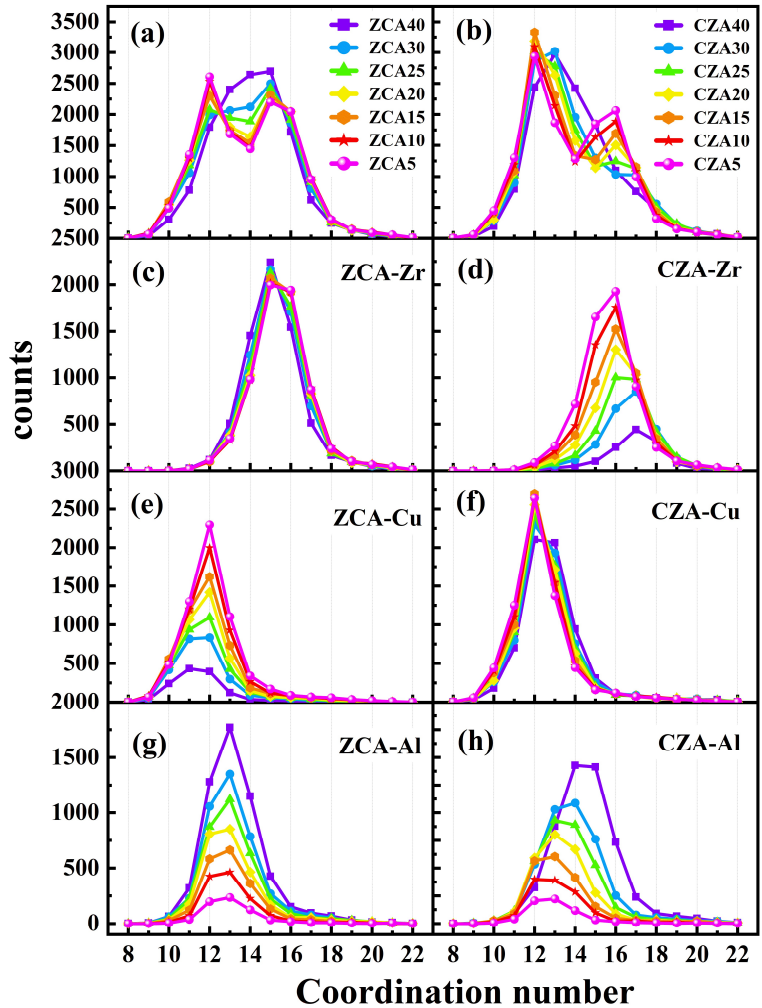

**Figure 7:** Distribution of atomic coordination numbers in ZCA and CZA metallic glasses. (a-b) average, (c-d) Zr, (e-f) Cu and (g-h) Al coordination numbers

WC-CSRO parameter, $\alpha_{ij}$, (Eq. 2) gives a quantitative measure of the chemical ordering and the strength of different atomic interactions. Fig. 8(a) and 8(b) show $\alpha_{ij}$ for different atomic pairs in ZCA and CZA, respectively. Negative values of $\alpha_{ij}$ for Zr-Al and Al-Zr pairs in the two alloy types for all the compositions are manifestations of the preference of Zr and Al atoms to be each other's neighbors. It is consistent with the highest $\Delta H_{ij}^{mix}$ for Zr-Al (-44 kJ/mol). $\Delta H_{ij}^{mix}$ for Cu-Zr and Cu-Al are -23 kJ/mol and -1 kJ/mol, respectively. So, the negative $\alpha_{ij}$ for Cu-Zr pairs, which suggest a preference of Cu atoms for Zr atoms as neighbours, is understandable. The positive $\alpha_{ij}$ for Zr-Cu indicate that Zr atoms avoids Cu neighbors in favor of Al atoms as $\Delta H_{ij}^{mix}$ for Zr-Al is nearly double the value for Zr-Cu. $\alpha_{ij}$ for Cu-Al and Al-Cu exhibit the most remarkable composition dependence in CZA glasses. The positive $\alpha_{ij}$ for Cu-Al decreases with the increase in Al concentration until Al% =15. Beyond that, it becomes negative and, increases in magnitude, which indicates an increasing tendency of Cu atoms to accommodate Al neighbors. Similar changes are observed for Al-Cu except that the positive $\alpha_{ij}$ becomes negative at Al% =20. Thus, Cu-Al CSRO could be a key factor in the composition dependence of the structural and thermodynamic properties of the ZCA and CZA metallic glasses.

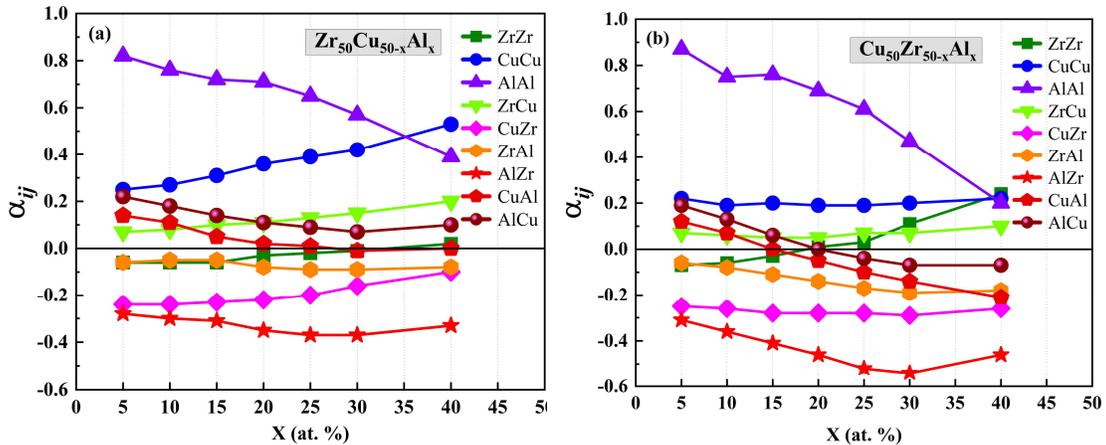

**Figure 8:** Warren-Cowley CSRO parameter $\alpha_{ij}$ for, (a) ZCA, (b) CZA metallic glasses

*3.4 Topological short- and medium-range ordering*

S(q) results in Sec. 3.2 point towards a higher degree of disorder in CZA than ZCA metallic glasses. The icosahedral short- and medium-range ordering is evident from the splitting of the second peak and the correlation of the peak positions with the first peak. [59] In general, it is well-established that local atomic clusters with five-fold symmetry, mainly icosahedral clusters, are the most prevalent short-range structural features in the metallic glasses and, they are closely linked to their dynamics and GFA. [19,20,36,37,40,60–64] The local structure of the metallic glasses is investigated using the Voronoi analysis (Sec. 2). The fractions of the most dominant icosahedra-like polyhedra in ZCA and CZA metallic glasses are shown in Fig. 9(a) and 9(b), respectively. Most of the 10–13 coordinated icosahedra-like polyhedra are Cu-centered and Al-centered. So, the population of these polyhedra gives an idea about the impact of the relative concentrations of Al and Cu in the ZCA and CZA alloys. Icosahedra, <0,0,12,0>, are energetically the most favoured local structures in materials with non-directional atomic interactions like metallic glasses. Although these icosahedra are geometrically imperfect, they are often termed full icosahedra (FI) due to the five-fold symmetry of their 12 faces. The atomic radii ratio and the heat of mixing are critical factors that make the formation of stable FIs sensitive to the concentration of different atom types in metallic glasses. The increasing Al% in Zr-rich ZCA glasses gives rise to higher negative $\Delta H^{Chem}$ due to an increase in dominant Zr-Al interactions (See Fig.2) and, should favour the formation of FIs. However, Fig. 8(a) suggests that the CSRO for unlike atom pairs is weaker. Therefore, FI formation would primarily depend on topological factors like atomic sizes. Accommodation of larger Zr and Al atoms among the 12 atoms of the first coordination cell of

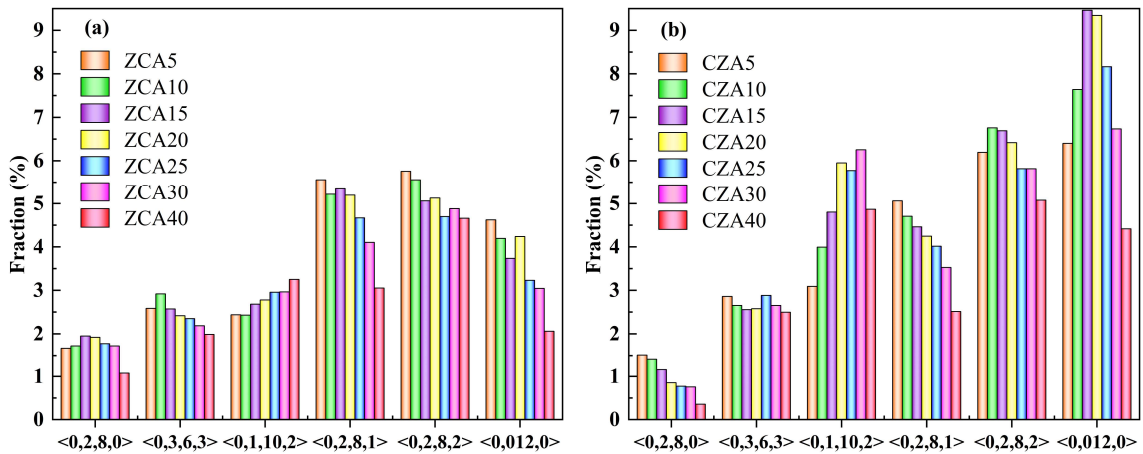

**Figure 9:** Population of dominant icosahedra-like Voronoi polyhedra in, (a) ZCA and, (b) CZA metallic glasses. The results are depicted on the same *y*-axis scale to signify the relative populations of each polyhedra type in the two metallic glasses.

Cu-centered and Al-centered FIs becomes increasingly difficult with the reducing concentration of Cu atoms in the ZCA glasses. So, the FI population in these glasses is significantly smaller than the defected icosahedra-like <0,2,8,2> polyhedra. Also, a nearly monotonic decrease in the fractions of FIs and <0,2,8,2> is observed with the increase in Al% in ZCA glasses. (Fig. 9(a))

In Cu-rich CZA glasses, high Al-Zr and Cu-Zr CSRO (Fig. 8(b)) and the Cu-Al bond shortening [22] promote Al-centered FI formation. So, it can be observed from Fig.8 that the fractions of FIs in these glasses are ~2 – 6 % higher than the ZCA glasses. A fast rise in the FI population is witnessed up to 15% Al in CZA glasses. Beyond that when Al% $\geq$ Zr%, that is Al% $\geq$ 25, a rapid fall in FI population is observed. This could be due to the reduction in the formation of Al-centered FIs as more Al atoms are accommodated in Cu-centered polyhedra (including FIs) and larger Al-centered polyhedra, which is further vindicated by the increasing Cu-Al and Al-Al CSRO for Al% > 25 (Fig. 8(b)). <0,1,10,2> is another polyhedra-type whose population is remarkably higher in Cu-rich CZA glasses than the Zr-rich ZCA glasses.

As the FIs and icosahedra-like polyhedra are known to form an interpenetrating backbone structure in ZrCu-based metallic glasses, [22,60,63,65,66] we investigate the degree of interconnectivity among the FIs. Two such polyhedra mainly connect through one of the four neighbour-sharing schemes: (1) vertex(1-atom) sharing, (2) edge (2-atom) sharing, (3) face (3-atom) sharing and, (4) pentagonal bi-cap (5-atom) sharing. Fig. 10(a) and 10(b) depict the degree of interconnectivity among FIs in the ZCA and CZA glasses, respectively. It can be observed that the degree of FI connectivity is remarkably higher in CZA glasses than in ZCA glasses. This is due to a significantly higher FI population in CZA glasses than in ZCA glasses. FIs are not the most dominant polyhedra in ZCA glasses. Rather, <0,2,8,2> and <0,2,8,1> are higher in population than FIs in these glasses. We can also see that the degrees of FI connectivity through the vertex, face and bi-cap sharing are nearly equal in CZA glasses, whereas the bi-cap sharing is higher than the vertex and face sharing in ZCA glasses. These results suggest a sparse spatial distribution of small clusters of connected FIs in ZCA glasses and a spatially heterogeneous distribution of large, compact clusters of highly interconnected FIs in CZA glasses. To validate this, we present the snapshots of the configurations of ZCA and CZA glasses in Fig. 10(c) and 10(d) where only the central atoms and the shared atoms of the interconnected FIs are shown. The trends in the composition dependence of FI connectivity in ZCA and CZA glasses are similar to those in FI populations. i.e. an increase in the FI population corresponds to an increase in FI connectivity and vice versa. As the icosahedral short- and medium-range ordering are closely linked to dynamics in the metallic glasses, we investigate the single particle dynamics in the following section.

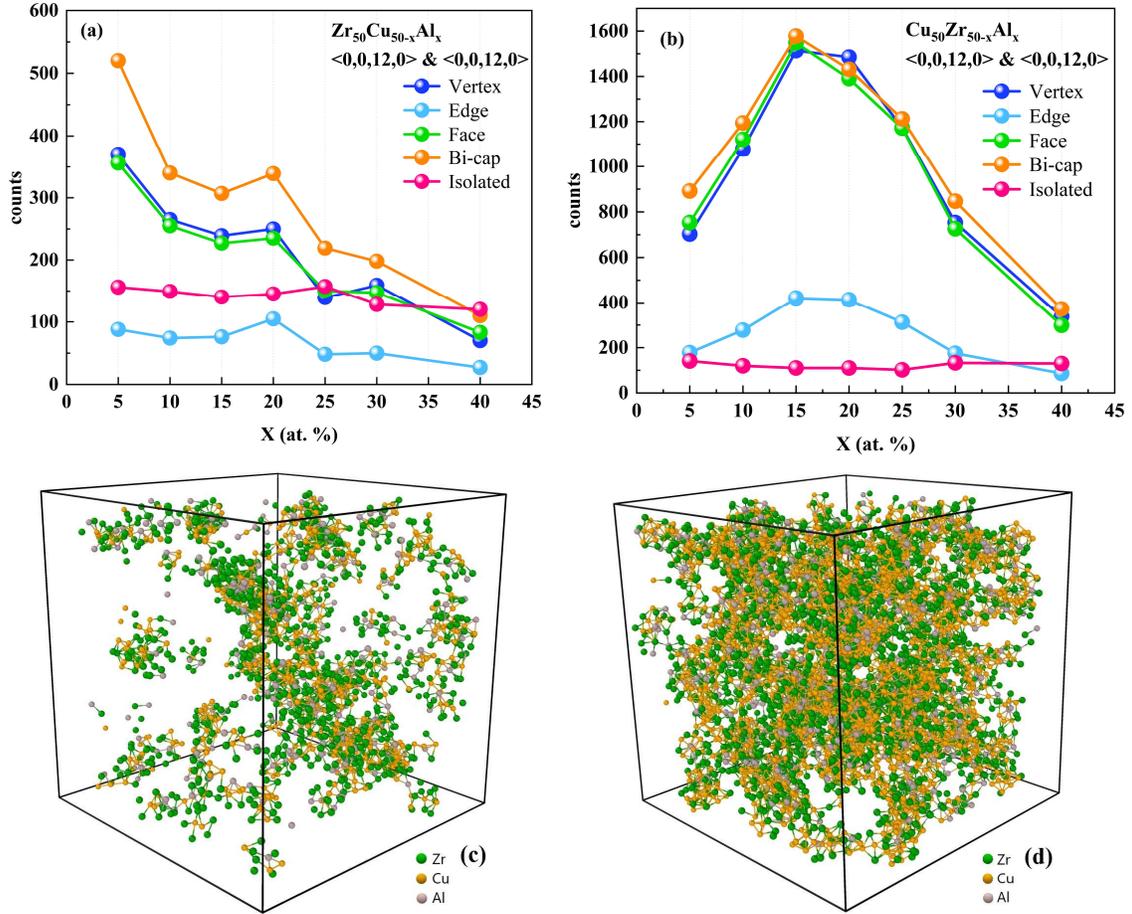

**Figure 10**: Interconnectivity among FIs in (a) ZCA and, (b) CZA glasses; Snapshots of a configuration of (c) ZCA and, (d) CZA metallic glasses with 15% Al. Only the central atoms and the shared atoms of the interconnected FIs are shown.

*3.5 Single particle dynamics: Structural relaxation, diffusion and dynamic heterogeneities*

A rapid non-Arrhenius slowdown of atomic dynamics during liquid-to-glass transition is a characteristic feature of metallic glass-forming liquids. According to the well-known Angell plot, [67] the degree of deviation of the temperature dependence of the structural relaxation time ($\tau_\alpha$) from the Arrhenius behaviour gives a measure of the GFA of the liquids. The rapid dynamics slowdown in Zr-Cu-Al alloys in the supercooled region has been attributed to the evolution of dynamic heterogeneities linked to the fast growth of short- and medium-range icosahedral ordering. [26,29,32,37,39,40]To understand the compositional effects on the dynamics of ZCA and CZA alloys, we look into the temperature dependence of $\tau_\alpha$, the self-diffusion coefficient ($D_s$) and the dynamic heterogeneities. $\tau_\alpha$ and $D_s$ are determined using the self-intermediate scattering function, $F_s(q_1, t)$, and the mean-square displacement, $\langle r^2(t) \rangle$, respectively through the analysis of the particle trajectories of equilibrated alloy configurations. $q_1$ corresponds to the position of the first(main) peak of S(q). The details and results of $F_s(q_1, t)$, and $\langle r^2(t) \rangle$ are given in the Supplementary Material.

As the relative concentrations of Zr and Cu are different in ZCA and CZA alloys, we consider the $\tau_\alpha$ and $D_s$ of Al atoms for a fair comparison of compositional effects in these alloys. The plots of the temperature dependence of $\tau_\alpha$ and $D_s$ for Al atoms in ZCA and CZA alloys are shown in Fig. 11 (a) and 11(b), respectively. The observed temperature dependence of $\tau_\alpha$ and $D_s$ is characteristic of fragile glass-formers, where an Arrhenius behaviour in the high temperature range above melting temperature crossover to a non-Arrhenius behaviour in the supercooled region. It can be observed from Fig.11(a) that at any given temperature in the investigated range 850 K – 2000 K, the values of $\tau_\alpha$ for ZCA alloys are very close and, hence, $\tau_\alpha$ curves nearly overlap. It is the same case for $D_s$ as well. Fig. 11(c) provides a better perspective where the compositional dependence of $\tau_\alpha$ in ZCA and CZA alloys at 850 K is shown. Thus, the kinetics of the relaxation and the diffusion processes during the liquid-to-glass transition in Zr-rich ZCA alloys, and hence GFA, remain unaffected by the large variations in Cu and Al concentrations. This explains the compositional invariance of $T_f$ in ZCA alloys observed in Fig. 1.

The temperature dependence of $\tau_\alpha$ and $D_s$ in CZA alloys with Al% ≤ 20 show very small compositional variations. It suggests similar relaxation and diffusion kinetics in these alloys. For the CZA alloys with Al% > 20, the temperature dependence of $\tau_\alpha$ and $D_s$ indicate increasingly faster atomic dynamics and lower GFA compared to the alloys with Al% ≤ 20. Considering the composition dependence of $\tau_\alpha$ at 850 K shown in Fig. 11(c), it can be said that CZA alloys with Al% ≤ 20 possess high kinetic stability and GFA. In contrast, the kinetic stability and the GFA rapidly deteriorate beyond Al% > 20. The rapid dynamics slowdown in metallic glass-forming liquids in the supercooled region has been attributed to the evolution of dynamic heterogeneities (DH) which refers to spatially heterogenous distribution of atomic mobilities. Therefore, it is pertinent to assess the correlation of DH with the observed compositional effects on the structural relaxation and diffusion in ZCA and CZA alloys.

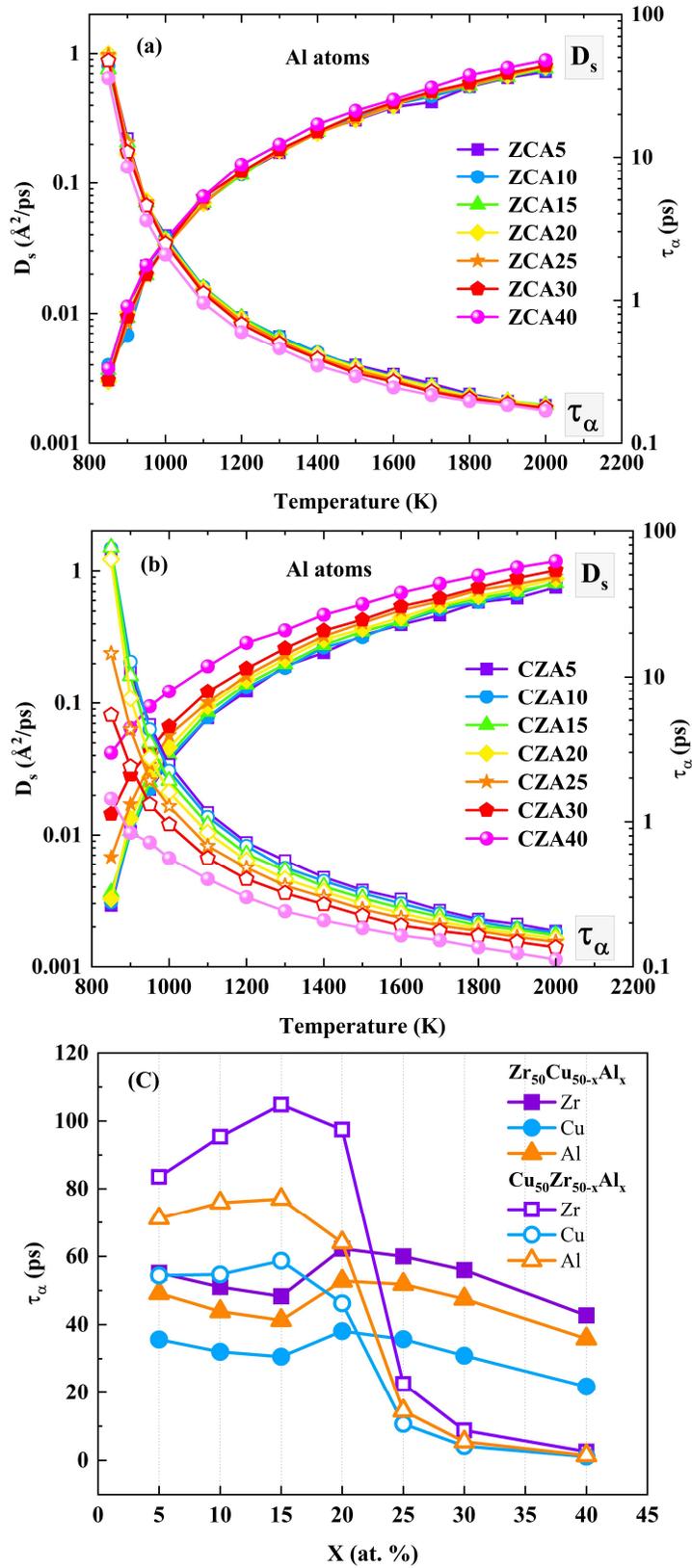

**Figure 11:** Temperature dependence of structural relaxation time ($\tau_\alpha$) and self-diffusion coefficient ($D_s$) in (a) ZCA and, (b) CZA alloys; (c) composition dependence of $\tau_\alpha$ at 850 K

The non-Gaussian parameter is a common measure of DH and, it is defined as, [68] $\alpha_2(t) = 3 \langle \sum_{j=1}^N |r_j(t) - r_j(0)|^4 \rangle / 5 \langle \sum_{j=1}^N |r_j(t) - r_j(0)|^2 \rangle^2 - 1$. It is a measure of the deviation of the distribution of atomic displacements from the Gaussian distribution. The results of elemental $\alpha_2(t)$ in ZCA and CZA alloys at 850 K are shown in Fig. 12. It can be observed from Fig.12(a) – (c) that a significant DH of Zr, Cu, Al atoms prevail in the ZCA alloys and, show marginal non-monotonic compositional variations except for the Al atoms in ZCA with 5% Al. In CZA alloys with Al% ≤ 20, predominant DH with small increase till 15% Al can observed in Fig. 12(d) – (f). Beyond Al% > 20, a drastic reduction in DH suggests that the atomic dynamics is increasingly becoming homogenous and faster. The compositional invariance of DH in ZCA alloys and the trends in the compositional variation of DH below and above 20% Al in CZA alloys are in exact correspondence with those observed for $\tau_\alpha$ and $D_s$.

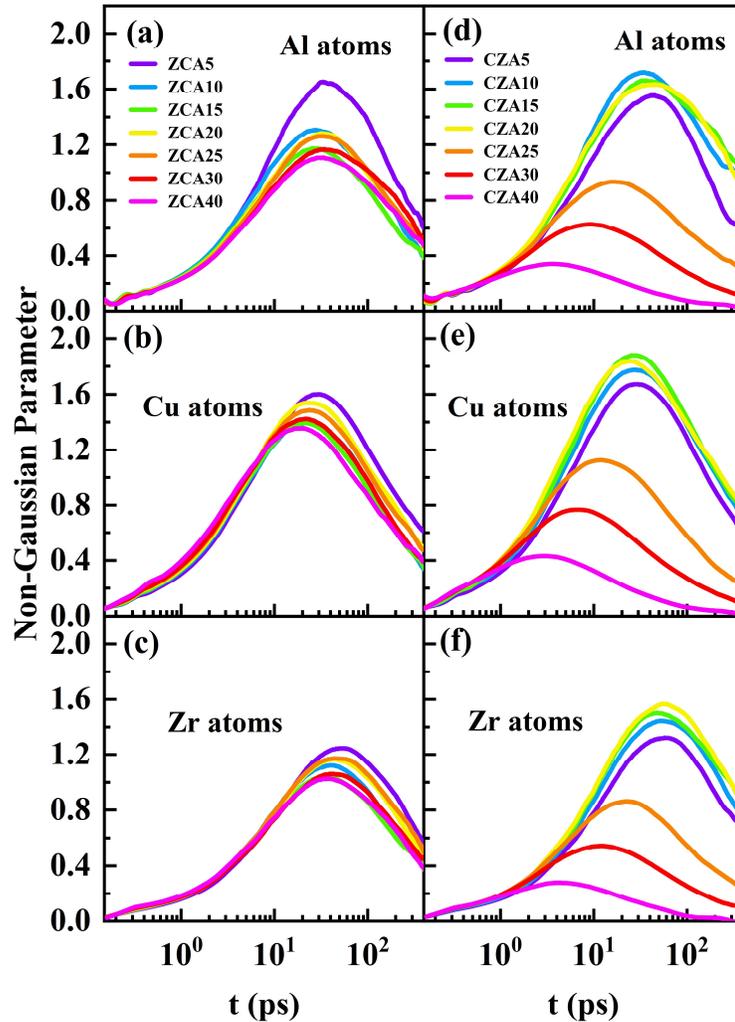

**Figure 12:** Elemental Non-Gaussian parameter in (a)-(c) ZCA and, (d)-(f) CZA alloys at 850 K

Thus, DH are directly correlated to the structural relaxation and diffusion in the Zr-Cu-Al alloys. DH are also shown to be linked to local structural heterogeneities that arise mainly due to short- and medium-range icosahedral ordering in metallic glass-forming liquids. [22,60,63,65,66] As icosahedral clusters are dynamically more stable, a larger degree of icosahedral short- and medium-range ordering with spatially heterogenous distribution gives rise to higher DH. A lower population of FIs (<0,0,12,0>) relative to other dominant icosahedra-like polyhedra (<0,2,8,1>, <0,2,8,2>) (Fig. 9(a)), a smaller degree of FI connectivity (Fig. 10(a) & (c)) and their marginal compositional dependence in ZCA alloys can be soundly related to the small compositional variation of DH seen in Fig. 12 (a) – (c). Similarly, a prominent non-monotonic compositional dependence of the FI population (Fig. 9(b)) and their large-scale interconnectivity (Fig. 10(b) & (d)) correspond to a remarkable non-monotonic compositional dependence in DH in CZA alloys as observed in Fig 12(d) – (f).

### 3.6 Mechanical Properties: Young's Modulus and Yield Stress

Structural heterogeneities and the associated DH in the metallic glasses are intimately linked to their plastic deformation and mechanical properties such as the elastic modulus and yield stress. [65,69–74] Local structural inhomogeneities that include densely-packed icosahedral clusters with a high degree of five-fold symmetry and loosely-packed atomic clusters devoid of significant five-fold symmetry are signatures of the strain localization and shear transformations leading to catastrophic failure in metallic glasses. [21,70] It has been shown that a higher population of FIs and higher degree of local five-fold symmetry (LFFS) are the primary factors inhibiting plastic deformation in Cu-Zr and Cu-Zr-Al metallic glasses. [21,70] So, our results of the topological short- and medium-range ordering (Sec. 3.4) and the DH (Sec.3.5) prompted the investigation of the plastic deformation in ZCA and CZA metallic glasses. We have analyzed the mechanical behavior of these glasses under tensile stresses at 300K. A uniaxial tensile stress is applied to the cubic simulation box in the z-direction with a strain rate of 0.2 $fs^{-1}$ in NPT ensemble. Periodic boundary conditions were applied in all the three directions. During the tensile deformation, the box length in z-direction increases periodically according to the applied strain rate, whereas the box lengths in x- and y-directions were allowed to adjust to maintain zero system pressure.

The stress-strain curves for the ZCA and CZA metallic glasses are shown in Fig. 13(a) and 13(b), respectively. The Young's modulus (E) and the yield stress determined from the curves are plotted in Fig. 13(c) and 13(d), respectively. The impact of icosahedral ordering, structural heterogeneity and the associated DH on the plastic deformation is evident from these results. FI population is significantly less than other icosahedra-like polyhedra ($<0,2,8,2>$,$<0,2,8,1>$) in ZCA alloys(Fig. 9(a)). The FIs are sparsely distributed with a lower degree of interconnectivity as can be seen from Fig. 10(a) and 10(c). It implies that local regions with lower degree of LFFS are more prevalent than the regions with high LFFS. Therefore, plastic deformation is easier in ZCA alloys. In CZA alloys, a predominant FI population (Fig. 9(b)) and a high degree of interconnection (Fig. 10(b)) give rise to large regions with high LFFS (Fig. 10(d)) hindering plastic deformation. So, the yield stress in CZA glasses is found to be higher than the ZCA glasses except the case of 5% Al. Moreover, a direct correlation between the FI population and the yield stress can be observed in both ZCA and CZA glasses. The FI population and the yield stress both show a remarkable non-monotonic change with the change in the Al concentration in CZA glasses. The yield stress is highest for CZA glass with 15% Al. A Bayesian exploration of the composition space of CuZrAl metallic glasses for mechanical properties demonstrates that the system has an optimal composition window for the yield stress around Al and Zr concentrations of 15 % and 30 %, respectively. [75] In ZCA glasses, even though the FIs are not the most dominant polyhedra, the change in its population directly corresponds to the change in the yield stress. A study of $Cu_{46}Zr_{46-x}Al_x$ ($x = 0,1,3,5,7$) glasses shows that an increase in the FI fraction makes the glass more resistant to plastic deformation. [69] A recent study of $Cu_{50}Zr_{50}$, $Cu_{47.5}Zr_{47.5}Al_5$, $Cu_{45}Zr_{45}Al_{10}$, and $Cu_{40}Zr_{40}Al_{20}$ glasses revealed that the addition of Al up to 20% significantly enhances plastic deformation by reducing shear band localization. [74] This study also highlights the significance of the local structural heterogeneity associated with FI evolution to the plastic deformation. Thus, our results reinforce the preposition by Cheng et al [69] that the fraction of FIs could be a good structural indicator for the plastic behaviour of the metallic glasses. Finally, it should be noted that Young's modulus exhibits marginal compositional variations up to 25% Al in ZCA and CZA glasses and, it is higher in CZA than ZCA. The higher E for CZA could be attributed to higher atomic packing and structural relaxation time due to significant icosahedral short- and medium-range ordering.

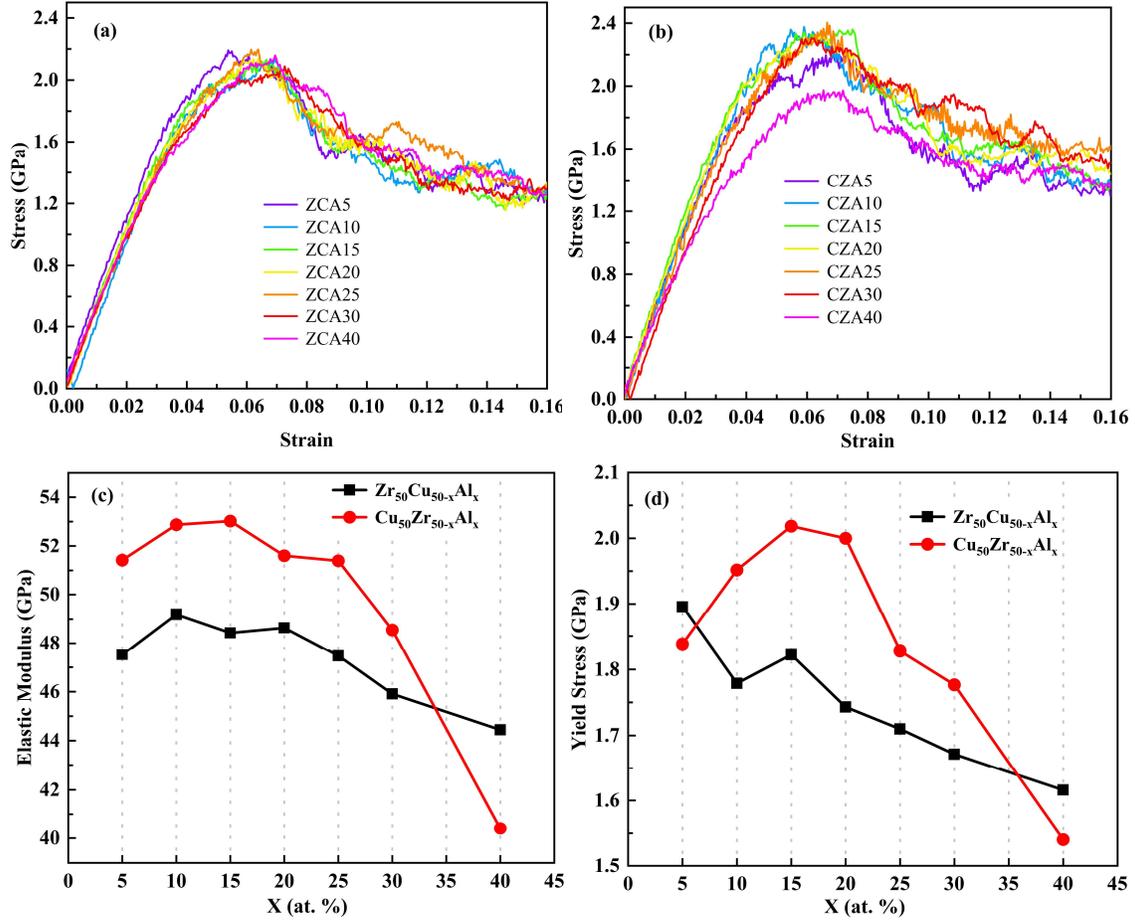

**Figure 13:** Stress-strain curves during tensile loading of (a) ZCA and , (b) CZA glasses at 300K; (c) Young's modulus, and (d) yield stress for ZCA and CZA glasses, respectively.

**Conclusions**

The findings of the present investigations of structure, dynamics, thermodynamic and mechanical properties of $Zr_{50}Cu_{50-x}Al_x$ (ZCA) and $Cu_{50}Zr_{50-x}Al_x$ (CZA) alloys (x=5,10,15,20,25,30,40), covering a wide compositional space, lead to some important conclusions that could serve as overarching guidelines for the choice of good glass-forming alloy compositions giving Zr-Cu-Al glasses with tailored thermal and mechanical properties.

Two strikingly different compositional dependence of the structure, dynamics, thermodynamic and mechanical properties of CZA and ZCA alloys are observed. The FWHM ($\Delta q$) of the first peak of S(q) and the temperature dependence of structural relaxation time ($\tau_\alpha$) in supercooled region, which are important structural and dynamical indicators, respectively for the GFA of the glass-forming alloys, show remarkable non-monotonic compositional dependence in CZA alloys. The $\Delta q$ and the temperature dependence of $\tau_\alpha$ in the CZA alloys combinedly indicate that the alloys with Al% ≤20 possess good GFA and, these glasses exhibit good plasticity with high yield stress and elastic modulus. In ZCA alloys, $\Delta q$ is significantly smaller than CZA glasses and the temperature dependence of $\tau_\alpha$ show very little compositional variation up to Al% ≤30. The results make it amply clear that the GFA of ZCA alloys is lower than the CZA alloys. The thermal (specific heat) and mechanical properties (yield stress, elastic modulus) of ZCA alloys with Al% ≤30 are significantly lower than CZA alloys and, show very small compositional changes. Overall, present results along with earlier studies [45,75] suggest that a good glass-forming Zr-Cu-Al alloy composition leading to a MG with good thermal and mechanical properties should be Cu-rich with Zr concentration in the window 30% – 35% and Al% ≤ 20.

Our results also highlight the impact of icosahedral short- and medium-range ordering on the dynamics and mechanical properties of the alloys. It is observed that the fractions of FIs and the degree of their interconnectivity are directly correlated to the structural relaxation, DH and the mechanical properties in ZCA and CZA alloys. So, notwithstanding the complex interplay of chemical-ordering and topological ordering in governing the glass-formation metallic alloys, FI population could still be a good preliminary indicator of the GFA and mechanical properties of the metallic alloys. Finally, we can make general remarks from the results of $T_f$, $\tau_\alpha$ and $D_s$; that the kinetics of structural relaxation and diffusion processes is least affected by the compositional changes in Zr-rich ZCA alloys whereas the kinetics of the dynamical processes is sensitive to the concentration of Zr and Al in the Cu-rich CZA alloys.

**Acknowledgements**

Authors acknowledge the financial support from SERB, New Delhi through major research project CRG/2018/001552. JPA and MBV acknowledge Govt. of Gujarat for SHODH fellowship. The authors also acknowledge the use of computational facilities at the Department of Physics, Sardar Patel University (SPU) through the DST-FIST program of the Department of Science and Technology, Govt. of India and the PARAM SHAVAK supercomputing facility established by Govt. of Gujarat at Sardar Patel University.

# Supplementary Material

## Compositional Effects on Structure, Dynamics, Thermodynamic and Mechanical Properties of Zr-Cu-Al alloys

**Kamal G. Soni**[*], Jayraj P. Anadani, Mitanshu B. Vahiya and Kirit N. Lad[*]

*Department of Physics, Sardar Patel University, Vallabh Vidyanagar-388120, Gujarat, INDIA*

## 1. Fictive temperature

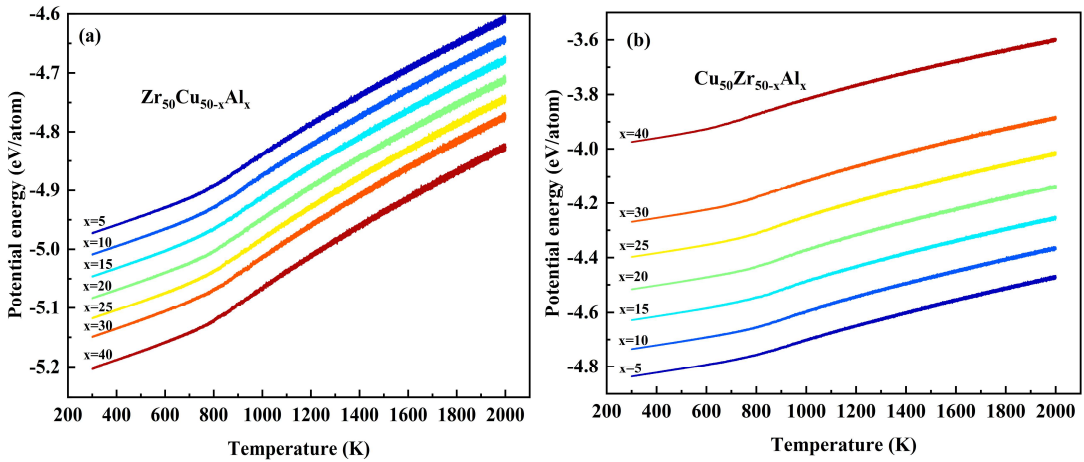

**Figure S1** Enthalpy vs. Temperature curves during quenching from 2000 K to 300 K for (a) $Zr_{50}Cu_{50-x}Al_x$ and (b) $Cu_{50}Zr_{50-x}Al_x$ systems with x = 5, 10, 15, 20, 25, 30, 40.

The fictive temperature ($T_f$) is determined from the enthalpy vs. temperature curves shown in Figure S1. The enthalpy values were obtained from molecular dynamics simulations under a constant cooling rate of 0.1 K/ps in NPT ensemble at zero pressure. An inflection point separating two distinct regions with different slopes indicates the glass transition. For each composition, linear fits were applied to the high-temperature supercooled liquid region and the low-temperature glassy region. The intersection point of these two lines was taken as the fictive temperature. This method provides a consistent estimate of $T_f$, which is used in the analysis of the thermal behavior of the metallic glass systems discussed in the main text.

## 2. Self-intermediate scattering function

The self-intermediate scattering function, $F_s(q_1, t)$, is a time correlation function used to probe the dynamical density correlations and structural relaxation of glass-forming systems. It is defined as

$$F_s(q,t) = \frac{1}{N}\left\langle \sum_{j=1}^{N} exp\{iq \cdot [r_j(t) - r_j(0)]\} \right\rangle$$

---

[*] Corresponding authors: knlad-phy@spuvvn.edu, kamalsoni@spuvvn.edu

The function describes how the positions of particles decorrelate over time at a specific length scale determined by q. For each composition, $F_s(q_1,t)$ was calculated at various temperatures in the range 850 K to 2000K. $q_1$ corresponds to the position of the first peak of S(q). The relaxation time ( $\tau_\alpha$ ) is considered to be the time when $F_s(q_1,t)$ decays to 1/e of its initial value. This method provides a quantitative measure of the structural relaxation behavior and allows for comparison of dynamical slowing down across compositions. We show only the relevant $F_s(q_1,t)$ results for Al atoms in ZCA and CZA atoms at 850 K.

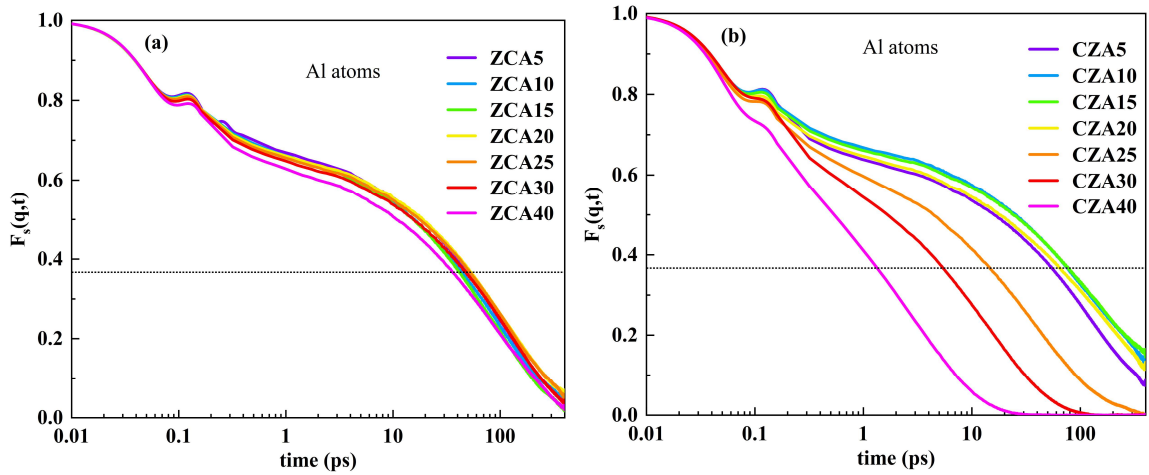

**Figure S2** Self-intermediate scattering function at 850 K for (a) $Zr_{50}Cu_{50-x}Al_x$ and, (b) $Cu_{50}Zr_{50-x}Al_x$ alloys with x = 5, 10, 15, 20, 25, 30, 40. The wave vector q corresponds to the position of the first peak in the static structure factor S(q) for each composition.

## 3. Mean Square Displacement

The mean square displacement is defined as

$$\langle \Delta r^2(t) \rangle = \frac{1}{N}\left\langle \sum_{j=1}^{N} |r_j(t) - r_j(0)|^2 \right\rangle$$

where N represents the number of atoms in a system, $r_j(0)$ and $r_j(t)$ are the initial and final coordinates of the $j^{th}$ atom, respectively. The <...> bracket denotes the ensemble average across all configurations. The self-diffusion coefficient can be calculated from the long-time limit ($t \to \infty$) of MSD as $D = \lim_{t\to\infty} \frac{1}{6t}\left\langle |r_j(t) - r_j(0)|^2 \right\rangle$. Representative results of MSD of Al atoms in the ZCA and CZA alloys at 850K are shown in Fig. S3(a) S3(b), respectively.

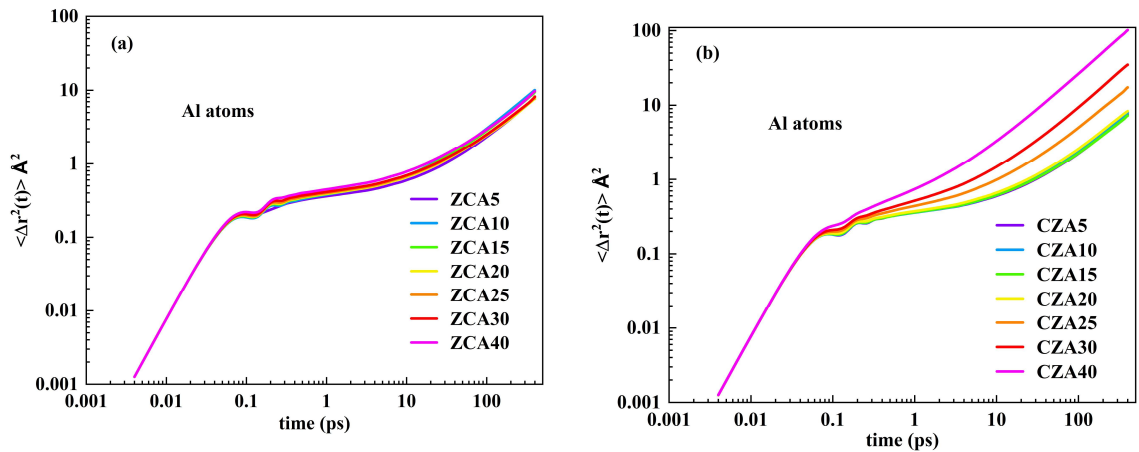

**Figure S3** Mean squared displacement (MSD) as a function of time at 850 K for (a) $Zr_{50}Cu_{50-x}Al_x$ and (b) $Cu_{50}Zr_{50-x}Al_x$ systems. The diffusion coefficient for each composition was calculated from the long-time linear regime of the MSD using the Einstein relation.